\documentclass[]{aa} 
\usepackage{natbib,graphicx}
\bibpunct{(}{)}{;}{a}{}{,}

\begin{document}

\title{A method to correct IACT data for atmospheric absorption due to
the Saharan Air Layer}

\author{
 D.~Dorner\inst {1} \and 
 K.~Nilsson\inst{2} \and
 T.~Bretz\inst{1}
}
 \institute {Universit\"at W\"urzburg, D-97074 W\"urzburg, Germany
 \and Tuorla Observatory, University of Turku, FI-21500 Piikki\"o, Finland}

\titlerunning{Method to correct for atmospheric absorption due to the Saharan Air Layer}

\date{Received 20.2.2008 / Accepted 1.8.2008 }

\offprints{Daniela Dorner \email{daniela.dorner@astro.uni-wuerzburg.de}}


\abstract{Using the atmosphere as a detector volume, Imaging Air
Cherenkov Telescopes (IACTs) depend highly on the properties and the
condition of the air mass above the telescope. On the Canary Island of
La Palma, where the Major Atmospheric Gamma-ray Imaging Cherenkov
telescope (MAGIC) is situated, the Saharan Air Layer (SAL) can cause
strong atmospheric absorption affecting the data quality and resulting
in a reduced gamma flux.}{To correlate IACT data with other
measurements, e.g. long-term monitoring or Multi-Wavelength (MWL)
studies, an accurate flux determination is mandatory. Therefore, a
method to correct the data for the effect of the SAL is
required.}{Three different measurements of the atmospheric absorption
are performed on La Palma. From the determined transmission, a
correction factor is calculated and applied to the MAGIC data.} {The
different transmission measurements from optical and IACT data provide
comparable results. MAGIC data of PG\,1553+113, taken during a MWL
campaign in July 2006, were analyzed using the presented method,
providing a corrected flux measurement for the study of the spectral
energy distribution of the source.}{}

\maketitle

\section{Introduction}
The Earth's atmosphere plays an important role in all ground-based
observations because it is traversed by light from all objects
observed. Scattering and absorption of light by the atmosphere is
variable and reduces the flux measured at the telescope.
On the Canary Island of La Palma, very strong atmospheric absorption
can occur, in case the meteorologic phenomenon Saharan Air Layer (SAL)
takes place. A warm air mass, consisting of mineral dust, travels from
the Sahara westward; this becomes the so-called SAL, when it is
undercut by a cool and moist air layer from the sea. Located between
1.5\,km and 5.5\,km above sea level (a.s.l.), the SAL, also known as
Calima, can extend above the North Atlantic Ocean as far as the United
States and the western Caribbean Sea \citep{sal}. 

Variable absorption can cause apparent flux variations, and the
absolute flux level is in all cases reduced. To to correct for this
reduction, the atmospheric absorption is measured. On La Palma, three
measurements are available. The Carlsberg Meridian Telescope (CMT)
provides nightly values of extinction. From the optical data of the KVA
and also from the MAGIC data itself, the atmospheric transmission can
be determined. The results of these measurements are discussed and
compared for the nights between July 15${\rm ^{th}}$ 2006 and July
28${\rm ^{th}}$ 2006.

A correction method was developed for the MAGIC telescope: From the
absorption measurements, a factor is calculated and used in the
analysis to correct for the effect of SAL.

\section{Feasibility Study}

Before the development of the method, the feasibility of a correction
based on the transmission measurement had to be checked. After
explaining the principle of the Imaging Air Cherenkov Technique, the
effect of the SAL on the Cherenkov data is explained, discussing in
particular the vertical shower development. 

\subsection{Imaging Air Cherenkov Technique}
\label{iact}
Particles enter the atmosphere, interact with nuclei and produce
electromagnetic or particle cascades. The Cherenkov light emitted by
these secondary particles traverses the remainder of the atmosphere and
is measured by a camera equipped with photon detectors. The morphology
of the recorded shower images is used to distinguish the images of
gamma-induced showers from those initiated by other particles, such as
protons. Therefore, the shower images are parameterized \citep{hillas},
and cuts in the corresponding distributions are used to suppress the
background \citep{cuts}. The energy spectrum is calculated by comparing
the parameters of the measured to those of the simulated data, for
which the energy is known. The most important parameter to reconstruct
the energy of a primary photon is SIZE, i.e.\ the total amount of light
in the shower image. Apart from minor dependences on other properties,
the energy is mainly proportional to the SIZE.

\subsection{Effect of Atmospheric Absorption}
The atmospheric absorption reduces the number of measured photons and
therefore the SIZE of the shower images. Consequently, the energy of
the primary particle is underestimated when comparing the SIZE of
affected data to that of showers simulated for an ideal atmosphere. For
an accurate energy reconstruction, the shower images have to be
corrected for the loss of light due to absorption, or the atmospheric
absorption has to be included in the simulation. As different amounts
of atmospheric absorption in the single nights would require different
sets of simulated data, a correction method for the data is preferred.

\subsection{Dependence on the Height in the Atmosphere}
\begin{figure*}
\sidecaption
  \includegraphics[width=12cm]{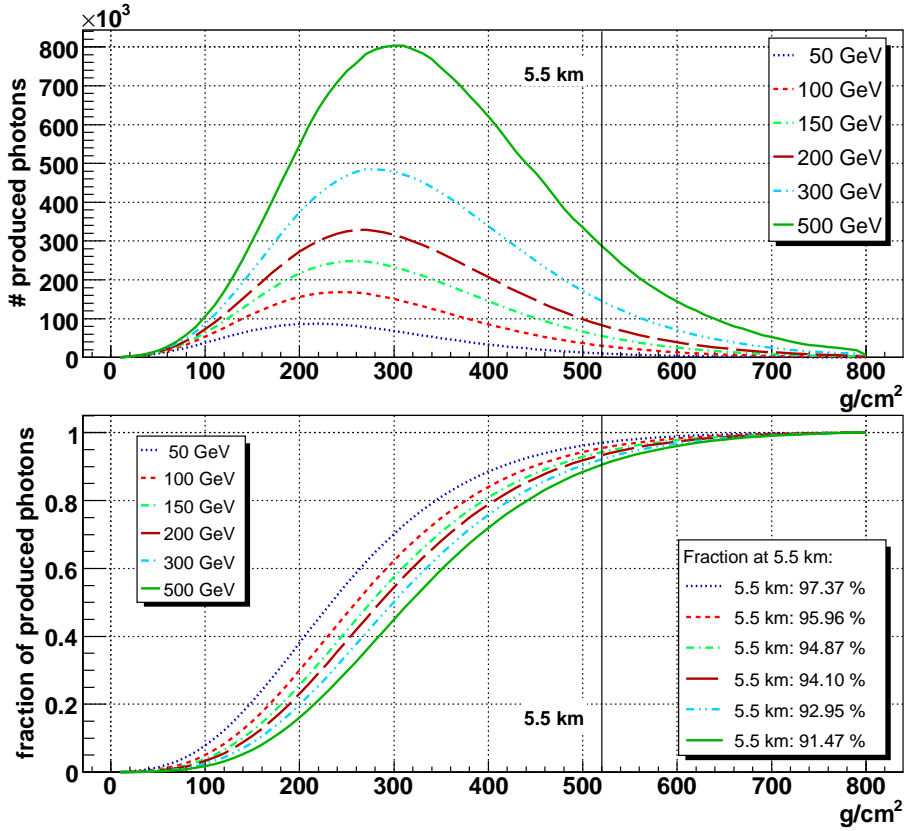}
\caption {In the upper plot, the number of produced photons (generated
with CORSIKA \citep{corsika}) is shown versus height (airmass in
$g/cm^2$) for monoenergetic showers with energies between 50~GeV and
500~GeV in different colors. Below, the integral fraction of produced
photons is plotted versus height for the same showers. In addition, the
integral fraction of light at to 5.5~km, indicating the upper border of
the Saharan Air Layer, is listed for each graph. \label{mclight}} 

\end{figure*}

Based on optical data, the values for transmission provided by CMT and
KVA correspond to light passing the whole atmosphere. The Cherenkov
light of a shower originates in a height between 20~km a.s.l. and the
ground. While the first interaction occurs at about 20~km a.s.l.,
the shower maximum is at around 10~km a.s.l. Figure~\ref{mclight} shows
the number of produced Cherenkov photons (upper plot) and the fraction
of produced light (lower plot) versus height. From the integral
fraction of light (lower plot), it can be concluded that, for primary
particles with an energy below 500\,GeV, more than 90\,\% of the shower
light is produced above 5.5\,km a.s.l. Consequently, less than 10\,\%
of the overall emitted light is absorbed by a slightly lower amount,
because it does not traverse the entire SAL. To a first-order
approximation, this small effect can be neglected, implying that a
factor calculated from the measured atmospheric transmission can be
used to correct the affected data.

\section{Measurements of the Atmospheric Absorption}
\label{measurements}

Absorption measurements are available from three different instruments
at the Observatorio del Roque de los Muchachos (ORM).

\subsection{CMT - Extinction Measurements}
The CMT \citep{cmt} is one of the optical telescopes in the ORM. Since
1984, it provides a value for the atmospheric extinction for each
night. Since April 1999, the measurement is completed in the r'-band at
an effective wavelength of 625\,nm \citep{cmtweb}. For a dust-free
night, the extinction for the passband r' was determined to be
0.09\,mag. The atmospheric transmission $\frac{I}{I_0}$ can be
calculated with the following formula, where $m_0$ is the extinction in
a dust-free night and $m$ the observed one:

\begin{equation}
\frac{I}{I_0} = 10^{\frac{m-m_0}{-2.5}}
\label{magnitude}
\end{equation}

While the Cherenkov spectrum extends between 280\,nm and 700\,nm and
peaks at about 320\,nm, the optical measurements are above 600\,nm.
Since the effect of SAL is dominated by Mie scattering, which is
independent of wavelength, the measured absorption can be used to
complete the correction.

Since the values of the CMT are provided at zenith, i.e.\ airmass
$X=1$, they must be corrected for the higher airmass when using them
for data of higher zenith distance. For the IACT data presented here,
the zenith distance is between 18$^\circ$ and 35$^\circ$ ($1.05 < X <
1.22$), which corresponds to a difference for the atmospheric
transmission of less than 5\,\% for the range of extinction values
shown here.

\subsection{KVA - Optical Data}
The KVA telescope\footnote{http://tur3.tur.iac.es/} consists of two
telescopes with 60\,cm and 35\,cm apertures. The latter is used for
photometric observations of TeV blazars in the R-band (effective
wavelength 640\,nm). The flux of the blazar is determined by measuring
the counts inside a circular aperture and comparing this to several
calibrated comparison stars in the field of the blazar, measured with
the same aperture radius. In this way, any influence of atmospheric
extinction is effectively eliminated. However, the comparison stars can
also be used to estimate the atmospheric extinction during the night.
In principle, the extinction could be determined by fitting a straight
line to an airmass--magnitude plot; the photometric accuracy of the
comparison star data in the literature (0.02\,mag-0.05\,mag) is,
however, insufficient for this. Only stars observed at low airmass
(X$<$1.2) were instead selected and, for each star, the zero point
value $z_0 = R_{instr} - R_{cal}$ was determined individually, where
R$_{instr}$ was the measured magnitude (scaled to an exposure time of
one second) and R$_{cal}$ was the calibrated R-band magnitude.
Averaging $z_0$ for each night, a value of 5.16 was found to provide a
good approximation of a photometric night. During nights affected by
the SAL, $z_0$ was higher. The transmission relative to a dust-free
night was computed by Eq. \ref{magnitude}, where $m_0$ was 5.16 and $m$
was taken to be the average $z_0$ for the night.

\subsection{MAGIC Data}
\label{cherdata}

In addition to optical measurements, the atmospheric transmission can
be determined directly from the IACT data.

A simple way is to use the absolute light yield measure obtained from
muon images as described in \citet{muoncal}. This method is based on
the fact that the spectral shape and flux of muons at the telescope are
well known. For a well-determined muon sample, only showers where the
telescope is inside the Cherenkov light cone are selected. As these
showers produce ring-like images in the camera, the background can be
completely suppressed. Comparing the measured light yield distribution
to that of a corresponding simulated sample, a ratio is determined that
can be used for an absolute light calibration to correct for changes in
the light efficiency of the telescope, e.g.\ mirror degradation. Since
this correction is completed on a monthly basis, short time variations
in the atmosphere are not taken into account. Consequently, the effect
of SAL is still evident in the measured muons. Having an impact
parameter smaller than the diameter of the mirror (17\,m), the selected
muons produce their light between 700\,m above the telescope and the
ground, i.e.\ between 2900\,m a.s.l.\ and \ 2200\,m a.s.l.\ in case of
the MAGIC telescope. To account for the fact that this light passes
only part of the SAL, the determined ratio is multiplied by a factor of
ten to estimate the absorption, assuming an effective production height
of 350\,m for Cherenkov light. Without any additional measurements, the
layering of the SAL is unknown. Accordingly, the results from the muons
are not very precise and should therefore not be used for correction,
but can serve as an independent crosscheck and first-order
approximation.

The absorption can instead be determined directly from the images of
the showers themselves by comparing the SIZE distributions of the
affected data with those of a dust-free night. The transmission is
calculated from the shift in the peak of these distributions, where $p$
is the position of the peak either for a dust-free night
($p_{dustfree}$) or for a night affected by the SAL ($p_{affected}$): 

\begin{equation}
\frac{I}{I_0} = \frac{p_{dustfree}-p_{affected}}{p_{dustfree}}
\label{sizecalc}
\end{equation}

For each night of affected data, one value $\frac{I}{I_0}$ was
determined. Since the observations used have almost the same zenith
angle distributions and the zenith distance is only between 18$^\circ$
and 35$^\circ$, the dependence on the zenith angle can be neglected. 

A similar method was studied for the Whipple telescope \citep{cosmic}
using the measured cosmic ray background for a relative calibration of
the detector response.

\section{Correction Method}
\label{method} In the analysis of MAGIC, there are several levels at
which possible corrections can be applied:

\begin{enumerate}
\item the factor in the absolute light calibration can be adapted to
correct for the additional absorption.  
\item the SIZE of the shower can be increased according to the
absorption
\item the correction can be included in the energy reconstruction.
\end{enumerate}

The last two possibilities obviously do not correct for changes in the
shower morphology. Losses of light at the shower margin might change
the shape of the image after cleaning. Consequently, common
background-suppression cuts, normally applied to all data, will give
less accurate results, since the morphology of the showers is different
from the one assumed by the cuts. Inter-night changes in the
atmospheric absorption complicate adapting the cuts to the changing
shower shape. In contrast, a correction applied before cleaning, i.e.
to the raw signal of all pixels, also correctly recovers the image
shape and parameters accordingly.

For this approach, an additional factor is applied in the conversion
from FADC counts to photo-electrons (phe). The correction is applied
according to the following formula: 

\begin{equation}
{\rm signal [phe]} = {\rm signal [FADC counts]} \cdot c_{cal} \cdot \frac{I_0}{I}
\label{correction}
\end{equation}

Applying this correction, obviously non-detected showers cannot be
recovered, but the SIZE distributions are not affected
significantly for an atmospheric absorption up to 40\,\%. For simulated
data (including absorption), the SIZE distributions agree well after
applying the correction for an absorption smaller than 40\,\%. In cases
of higher absorption, a more detailed Monte Carlo study is required,
and a more complex correction might be considered. 

Furthermore, the fluctuations in signals, such as those from night-sky
background light, increase. Comparing corrected dark-time observations
with data taken in strong moonlight \citep{moon}, the fluctuations
after correction are found to remain well below those for extreme light
conditions. By including arrival-time information in the image
cleaning, it could also be shown \citep{time} that even data with
strong fluctuations are properly handled. Hence, no further correction
is necessary to account for this effect if the correction is applied to
data taken under normal light conditions. Consequently, the calculated
image parameters are not skewed by the increase in the amplitude of the
fluctuations and therefore do not influence the efficiencies obtained
from simulated showers.

\section{Results}
To check the method, a source of constant flux is ideally used.
However, no affected MAGIC data of a steady source are available. Data
affected by the SAL are difficult to schedule, since these rare weather
conditions are hardly predictable and difficult to identify for the
support astronomers.

In July 2006, the MAGIC telescope observed the Active Galactic Nucleus
PG\,1553+113 contemporaneously with the X-ray satellite Suzaku, the
IACT array H.E.S.S.\ and the optical telescope KVA during a MWL
campaign.  Due to the SAL, the absorption for the single nights from
July 15${\rm ^{th}}$ to July 28${\rm ^{th}}$ varied between 5\,\% and
30\,\% according to the SIZE measurement. Therefore, the discussed
correction was applied to the MAGIC data.

\subsection{Measurements of the Atmospheric Transmission}
\begin{figure*}
\centering
\includegraphics[width=17cm]{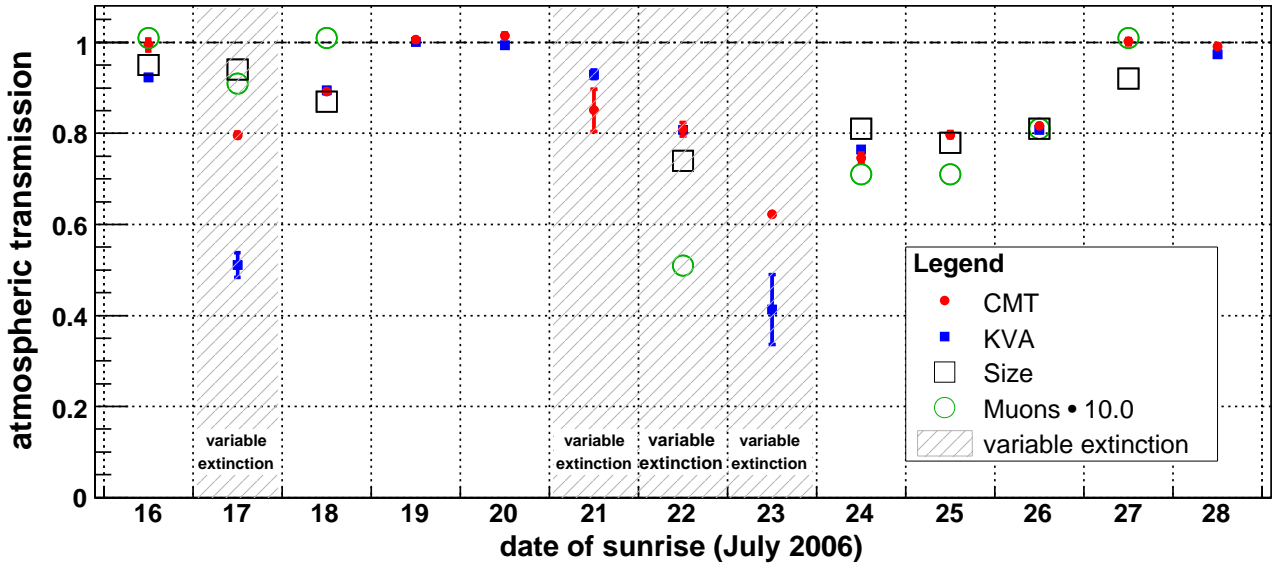}
\caption{ Atmospheric Transmission for the nights between July 16th and
28th 2006 from the measurements of the CMT (red dots), the KVA (blue
full squares) and MAGIC. From the latter, not only the SIZE measurement
(black open squares) is shown, but also the values derived from muon
rings (green circles) where the value is scaled by a factor ten to
account for the smaller absorption for muons. Nights with unstable
extinction are marked with gray shaded background. \label{extinction}} 

\end{figure*}

With the atmospheric transmission $\frac{I}{I_0}$ determined from the
presented measurements, comparable results were achieved for the eight
nights of observation of PG\,1553+113 in July 2006
(Fig.~\ref{extinction}). From the MAGIC data, the transmission was
determined only for the nights in which data of PG\,1553+113 were
available. During some nights, the atmospheric transmission was
changing, as seen both by the CMT and KVA. The KVA reports considerable
changes even within a few minutes. Consequently, the different
measurements agree less than for nights with stable extinction, since
they were probably acquired not precisely simultaneously. Unstable
conditions during the observations worsened the accuracy of the
correction. The remaining differences between the values from the KVA
and the CMT can be explained by the fact that the intrinsic errors in
magnitude for the comparison stars are larger in the case of KVA.
Deviations of the SIZE may come from the limited accuracy of the
method. For the muon measurements, the systematic uncertainties in the
values are larger due to the strong dependence on the vertical
distribution of the sand-dust in the SAL. Apart from small deviations
and the differences in the nights with variable extinction, the
measurements agree well. 

Since the CMT measurements are the most accurate, they were used for
the calculation of the correction factors.

\subsection{SIZE Distributions}
\begin{figure*}
\centering
  \includegraphics[width=17cm]{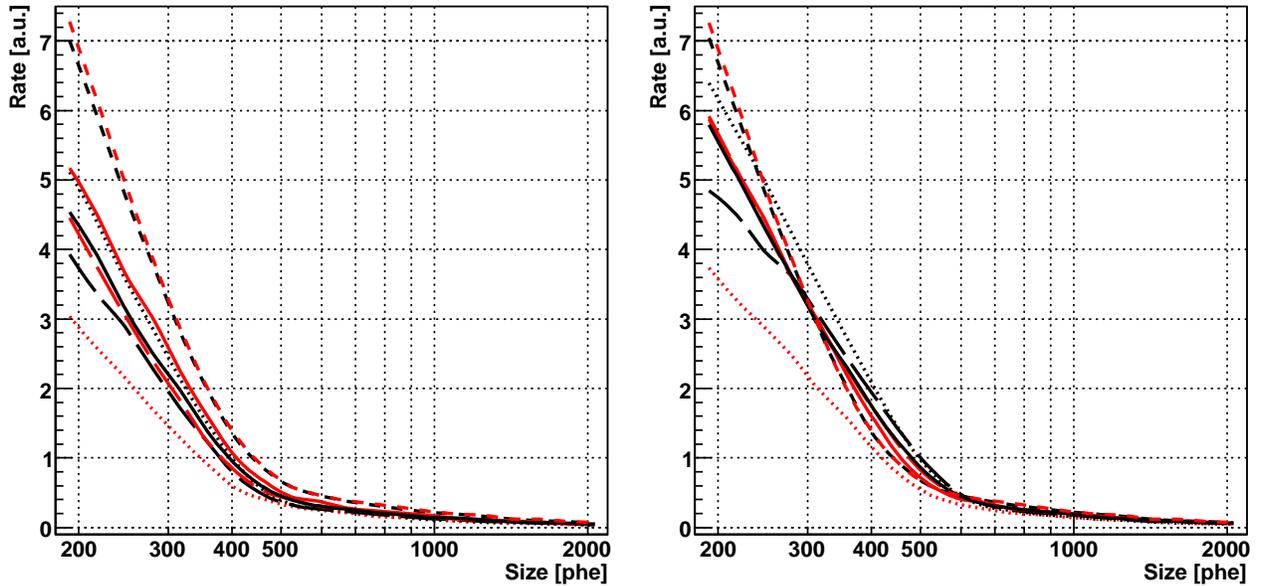}
\caption { SIZE distributions between 200\,phe and 2000\,phe (i.e.\
approximately from 100\,GeV to 1\,TeV) before (left) and after (right)
correcting for the effect of SAL. Each line corresponds to one
night and is scaled by the observation time. The color and style of
each line is different to distinguish the nights from each other.
\label{sizedistr}}

\end{figure*}

As explained in Sect.~\ref{iact}, the atmospheric extinction influences
mainly the SIZE of the showers. Figure~\ref{sizedistr} shows the
SIZE distributions for the range used for the analysis before (left)
and after (right) applying the correction with the values from the CMT.
For the shown distributions, special cuts were applied to remove
non-physical and muon events. Each curve corresponds to data of one
night. 

Before the correction, the SIZE distributions spread between 2.8\,a.u.\
and 7\,a.u.\ at 200\,phe. After the correction, the spread was reduced
to the interval between 4.7\,a.u.\ and 7\,a.u.\ apart from one outlier,
corresponding to one night of variable extinction (red dotted line,
22.7.2006). Apart from this and a second curve (black dotted line,
17.7.2006) with a deviation at around 300\,phe, the SIZE distributions
agree well after the correction for SIZE\,\textgreater\,200\,phe, i.e.\
in the range used in the analysis.

\subsection{Signal Reconstruction} 
The cuts for the background suppression mentioned in Sect.~\ref{iact}
include an intrinsic lower cut in SIZE, i.e.\ events with small SIZE
are removed. Consequently, events with reduced SIZEs due to atmospheric
absorption are cut away, resulting in a lower significance of the
signal if the same standard cuts are applied.

For PG\,1553+113, a significant number of events are expected at low
energies  due to the measured steep spectrum of the source. The
expected effect of reduced significance due to the SAL is observed for
data taken in July 2006: Without correction, a signal is obtained with
3.2\,$\sigma$ significance, improving to 5.0\,$\sigma$ with correction,
using identical cuts for the background suppression in both cases.

\subsection{Flux Determination}
Applying the correction as explained in Sect.~\ref{method}, the
8.5~hours of MAGIC data of PG\,1553+113 from July 2006 were analyzed.
Fitting a power law to the differential spectrum, a flux of
$(1.4\pm0.3)\cdot10^{-6}~{\rm ph\,TeV^{-1}s^{-1}m^{-2}}$ at 200\,GeV
and a spectral index of $-4.1\pm0.3$ were determined. A daily light
curve shows a flux consistent with a constant flux within the errors.
The results were shown in detail in \citet{magic1553-2} and were used
for MWL studies by \citet{mwl}.

Correcting the flux for the effect of the SAL, the systematic
uncertainty in the method of 5\,\% results in an additional error of
$0.2\cdot10^{-6}~{\rm ph\,TeV^{-1}s^{-1}m^{-2}}$ in the absolute flux.

\section{Conclusion and Outlook}

The different measurements of the atmospheric transmission from stellar
photometry (KVA, CMT) and IACT data (MAGIC) agree well for nights with
stable extinction. Since more than 90\,\% of the shower light
originates above the SAL, the measurements can be used directly to
correct the IACT data. By applying a correction factor in the
calibration, the data of PG\,1553+113 taken during the MWL campaign in
July 2006 for example were corrected successfully. This shows that the
loss of observation time due to atmospheric absorption can be reduced.

For additional checks, affected data of a steady source are required.
When such data become available within scheduled MAGIC observations,
the presented method can be tested further. 

The correction should not be applied blindly to data affected by
atmospheric absorption of another kind. For those, the vertical
distribution of the atmospheric conditions should be studied e.g.\ with
a Lidar (light detection and ranging), i.e.\ a device shooting a laser
beam into the atmosphere and measuring the backscattered light. The
possible use of information provided by a Lidar, however, still has to
be investigated. 

In cases of strong absorption, i.e.\ above 40\,\%, the method is
limited by the fluctuations in the night-sky background light and the
effect cannot be corrected completely.

\begin{acknowledgements}
We would like to thank the IAC for the excellent working conditions at
the Observatorio del Roque de los Muchachos in La Palma. The support of
the German BMBF is gratefully acknowledged. We thank Dafydd Wyn Evans
for his help and fruitful discussions about the data of the Carlsberg
Meridian Telescope and Dorota Sobzyncska for her help with the
simulated showers.  
\end{acknowledgements}

\bibliographystyle{aa}
\bibliography{9618}

\end{document}